\documentclass[%
 reprint,
 amsmath,amssymb,
 aps,
]{revtex4-2}

\usepackage{graphicx}% Include figure files
\usepackage{dcolumn}% Align table columns on decimal point
\usepackage{bm}% bold math
\usepackage{multirow}
\begin{document}

\preprint{APS/123-QED}

\title{Thermal and driven noise in Brillouin lasers}% Force line breaks with \\

\author{John H. Dallyn$^{1,4}$}
\email{jhd64@nau.edu}
\author{Kaikai Liu$^2$}
\author{Mark Harrington$^2$}
 \author{Grant Brodnik$^2$}
 \author{Peter T. Rakich$^3$}
 \author{Daniel J. Blumenthal$^2$}
\author{Ryan O. Behunin$^{1,4}$}
\email{ryan.behunin@nau.edu}
\affiliation{%
 $^1$Department of Applied Physics and Material Sciences, Northern Arizona University, Flagstaff, Arizona, USA
}%
\affiliation{%
 $^2$Department of Electrical and Computer Engineering, University of California Santa Barbara, Santa Barbara, CA, USA 
}%
\affiliation{%
 $^3$Department of Applied Physics, Yale University, New Haven, CT, USA 
}%
\affiliation{%
$^4$Center for Materials Interfaces in Research and Applications (¡MIRA!), Flagstaff, AZ
}%

\date{\today}% It is always \today, today,

%%%%%%%%%%%%%%%%%%%%%%%%%%%%%%%%%%%%%%%%%%%%%%%
\begin{abstract}
%%%%%%%%%%%%%%%%%%%%%%%%%%%%%%%%%%%%%%%%%%%%%%%
Owing to their highly coherent emission and compact form factor, Brillouin lasers have been identified as a valuable asset for applications including portable atomic clocks, precision sensors, coherent microwave synthesis and energy-efficient approaches to coherent communications. While the fundamental emission linewidth of these lasers can be very narrow, noise within dielectric materials leads to drift in the carrier frequency, posing vexing challenges for applications requiring ultra-stable emission. 
A unified understanding of Brillouin laser performance may provide critical insights to reach new levels of frequency stability, however existing noise models focus on only one or a few key noise sources, and do not capture the thermo-optic drift in the laser frequency produced by thermal fluctuations or absorbed power.
Here, we develop a coupled mode theory of Brillouin laser dynamics that accounts for dominant forms of noise in non-crystalline systems, capturing the salient features of the frequency and intensity noise for a variety of systems.  
As a result, theory and experiment can be directly compared to  identify key sources of noise and the frequency bands they impact, revealing strategies to improve the performance of Brillouin lasers and pave the way for highly-coherent sources of light on a chip. 
\end{abstract}

\maketitle

%\tableofcontents

%%%%%%%%%%%%%%%%%%%%%%%%%%%%%%%%%%%%%%%%%%%%%%%
\section{\label{sec:Intro}Introduction}
%%%%%%%%%%%%%%%%%%%%%%%%%%%%%%%%%%%%%%%%%%%%%%%
In recent years, the highly coherent emission produced by Brillouin lasers have enabled applications including compact atomic clocks  \cite{loh2020operation}, visible light sources \cite{chauhan2021visible}, precision gyroscopes \cite{zarinetchi1991stimulated,li2017microresonator,gundavarapu2019sub}, ultra-stable microwave generators \cite{li2013microwave,gundavarapu2019sub} and energy-efficient architectures for optical communications \cite{brodnik2021optically}. Key to these applications is the remarkable properties of Brillouin lasers that enable narrow fundamental emission linewidths \cite{smith1991narrow,li2012characterization,geng2006highly,grudinin2009brillouin,li2013microwave,gundavarapu2019sub,li2014low}. While these fundamental linewidths can reach sub-Hz levels \cite{li2012characterization,suh2017phonon,gundavarapu2019sub,li2013microwave,li2014low}, thermal and driven sources of noise lead frequency instability that produce drift in the laser emission  \cite{gorodetsky2004fundamental,savchenkov2007whispering,matsko2007whispering}.
Predictive models capturing all key noise sources may provide the insights to overcome these challenges and pave the way to the Hz-level frequency stability desired for applications such as precision time-keeping and spectroscopy.
However, empirically based models capturing the key features of Brillouin laser performance have not been developed. 

To reach the levels of performance demanded by precision applications for chip-scale Brillouin lasers, the noise spectra of these lasers must be understood and predicted.
Many facets of Brillouin laser performance have been described by models of transferred and fundamental noise \cite{debut2000linewidth,li2012characterization,behunin2018fundamental,loh2015noise,matsko2012stability}. Coupled mode and envelope theories predict phase and relative intensity noise transferred to the laser emission from the pump \cite{debut2000linewidth,li2012characterization,loh2015noise}, and have been used to derive a Schawlow-Townes-like linewidth that results from noise inherent to Brillouin scattering (i.e., the analog of spontaneous emission in the gain medium) \cite{li2012characterization}.  
However, existing models neglect photothermal and thermorefractive noise, which constitute fundamental sources of noise within dielectric resonators, and must be included to create predictive models. 
Investigations of the fundamental frequency stability of dielectric resonators lay the foundation for the understanding these effects as well as thermally driven sources of noise arising from Brownian, thermoelastic and pondermotive effects   \cite{gorodetsky2004fundamental,matsko2007whispering,savchenkov2007whispering,elshaari2016thermo,panuski2020fundamental}. 

In this paper, we develop a comprehensive model for Brillouin laser dynamics in non-crystalline dielectric resonators such as silica microresonators \cite{loh2015dual,loh2015noise,li2012characterization,li2013microwave,li2014low,lee2012chemically} and all-waveguide optical cavities \cite{kabakova2013narrow,eggleton2019brillouin,yang2018bridging,gundavarapu2019sub,chauhan2021visible,hu2014low}. 
By incorporating thermorefractive and photothermal noise, this model captures the noise sources critical to an understanding of the performance of on-chip Brillouin lasers \cite{matsko2007whispering,savchenkov2007whispering}. 
To validate our model, we compare the predicted noise spectra with measurements obtained from a Si$_3$N$_4$ photonic integrated Brillouin laser \cite{gundavarapu2019sub,chauhan2021visible}. Using experimentally derived parameters, these predictions capture the key features of the experimental laser spectra including frequency and intensity noise. By identifying and quantifying key sources of noise and the frequency bands they impact, these results reveal strategies to improve the performance of Brillouin lasers and pave the way for highly-coherent sources of light on a chip. 

The paper is organized as follows: Section II describes the physical origin of the dominant noise sources in non-crystalline Brillouin lasers. Section III outlines the coupled-mode theory capturing the dominant forms of noise. The laser's amplitude and phase dynamics are derived, including the impact of feedback that locks the pump frequency to the laser resonator.
Thermorefractive and photothermal noise are modelled by coupling the optical mode dynamics with the temperature field. We add temperature-dependent optical mode frequencies to the coupled mode dynamics and solve the stochastic dynamics of the temperature field that include drive terms associated with fundamental thermal fluctuations as well as the heat flux produced by the optical field. Section IV compares model predictions to measured frequency and intensity noise power spectra for an integrated waveguide Brillouin laser. This comparison, and quantitative agreement, enables the features of the noise spectra to be identified and understood, and therefore, through the equations given in Sec. III, to be controlled. Section V summarizes the key results of the paper. 

%%%%%%%%%%%%%%%%%%%%%%%%%%%%%%%%%%%%%%%%%%%%%%%
\section{Sources of Brillouin laser noise }
%%%%%%%%%%%%%%%%%%%%%%%%%%%%%%%%%%%%%%%%%%%%%%%
The performance of non-crystalline Brillouin lasers is determined by spontaneous emission (fundamental noise) \cite{li2012characterization}, thermal noise produced by both fundamental and driven thermal fluctuations, and noise transferred from the pump laser.  In this section, we describe the qualitative nature and physical origin of each of these noise sources. In later sections, we show how these noise sources are modeled and provide expressions for derived frequency and intensity noise power spectra.  

{\it Fundamental noise} is inherent to the laser amplification process. For a Brillouin laser, this noise is produced by spontaneous Stokes emission from thermally populated phonon modes. This effect has been explored in models utilizing coupled modes \cite{li2012characterization,behunin2018fundamental} and envelopes \cite{loh2015noise}, where  a Schawlow-Townes-like linewidth describes the frequency noise \cite{li2012characterization} and the intensity fluctuations exhibit relaxation oscillations \cite{loh2015noise,behunin2018fundamental}. 
   
{\it Thermorefractive noise} occurs when temperature fluctuations within a dielectric resonator lead to local shifts in the index of refraction through the thermo-optic effect. Consequently, a fluctuation in temperature can shift the resonant frequencies of a cavity \cite{gorodetsky2004fundamental}. Because this noise source scales with the inverse resonator mode volume, thermorefractive noise can be a critical form of frequency instability within microresonators.

{\it Photothermal noise} results from frequency fluctuations  originating from thermo-optic shifts in the index of refraction that are driven by optical absorption of fluctuating intracavity power. These power fluctuations arise from both fundamental amplitude noise of both pump and Stokes modes within the resonator as well as relative intensity noise (RIN)  present in the source laser used to drive Brillouin lasing.    

{\it Transferred pump noise} can be imprinted on the Brillouin laser emission. While SBS is known for producing laser emission with noise that is drastically compressed compared to the pump, this filtering ability is ultimately limited by relative decay rates of the optical and acoustic modes \cite{debut2000linewidth}. In addition, RIN of the pump also drives intensity fluctuations of SBS emission, impacting the power stability and leading to one source of photothermal noise.  

{\it Other noise sources} are present within Brillouin lasers and require careful analysis in systems, such as microtoroids or cystalline resonators, where radiation pressure and thermal expansion are significant compared to the thermo-optic effect. These pondermotive and thermoelastic effects as well as Brownian motion of the resonator structure are small compared to the noise sources described above within the non-crystalline SBS lasers considered here \cite{matsko2007whispering,savchenkov2007whispering}.  

%%%%%%%%%%%%%%%%%%%%%%%%%%%%%%%%%%%%%%%%%%%%%%%
\section{Theory}
%%%%%%%%%%%%%%%%%%%%%%%%%%%%%%%%%%%%%%%%%%%%%%%

To model Brillouin laser dynamics, we use a coupled-mode theory, treating the optical and  acoustic modes as mean-field (lumped) elements \cite{li2012characterization,behunin2018fundamental}. The validity of this model requires that the temporal changes in the electric field amplitude, the loaded optical decay rate and the gain bandwidth are all much smaller than the free spectral range and relevant resonance frequencies of the resonator, i.e., intermodal scattering is negligible and the rotating wave approximation (RWA) is valid. %(e.g. see Ref. \cite{debut2000linewidth} to see how the coupled-envelope picture reduces to coupled-modes). 
Owing to the relatively large strength of electrostriction, we neglect effects produced by the Kerr nonlinearity, i.e., self- and cross-phase modulation, valid for modest powers and relatively small refractive indices.

Under these assumptions, Brillouin laser dynamics can be modeled using the Hamiltonian $H = H_0 + H_{\rm int}$ where the $H_0$ decribes the uncoupled dynamics of the pump, Stokes and phonon modes 
\begin{equation}\label{Ham0}
    H_0=\hbar\omega_p a^{\dagger}_p a_p + \hbar\omega_S a^{\dagger}_S a_S + \hbar\Omega b^{\dagger} b
\end{equation}
and $H_{\rm int}$ quantifies the coupling of these modes through electrostriction
\begin{equation}\label{HamInt}
    H_{\rm int} =  \hbar g a^{\dagger}_p a_S b + \hbar g^* a_p a^{\dagger}_S b^{\dagger}
\end{equation}
where we neglect the effect of cascaded-order lasing. 
Here, $\omega_p$, $a_p$, $a_p^\dag$ and  ($\omega_S$, $a_S$ and $a_S^\dag$) denote the pump (Stokes) frequency and annihilation and creation operators, respectively, while  $\Omega$, $b$ and $b^\dag$ are the phonon mode frequency and annihilation and creation operators. The parameter $g$ is a coupling rate quantifying the electrostrictive interaction between the pump, Stokes and phonon modes which is determined from the spatial overlap of the acoustic and optical modes and the photoelastic tensor \cite{behunin2018fundamental}. 

Using this Hamiltonian and adding the effects of optical and acoustic dissipation as well as an external pump, we derive equations of motion for the SBS laser. Under the assumption that pump, Stokes and phonon modes are phased matched, i.e. $\omega_p = \omega_S + \Omega$, we find Heisenberg-Langevin equations of motion, evaluated in a rotating frame about the resonance frequency of each field, given by
\begin{equation}\label{apeom}
    \dot{a}_p=-\frac{1}{2}\gamma_p a_p - i g a_S b + \sqrt{\gamma_{ext}}F_{ext} + \eta_p
\end{equation}
\begin{equation}\label{aseom}
    \dot{a}_S=-\frac{1}{2}\gamma_S a_S - i g^* a_p b^\dagger + \eta_S 
\end{equation}
\begin{equation}\label{beom}
    \dot{b}=-\frac{1}{2}\Gamma b - i g^* a_p a_S^\dagger + \xi
\end{equation}
\newline
where $\Gamma$, $\gamma_p$ and $\gamma_S$ are the respective acoustic, pump and Stokes mode decay rates, and $\eta_p$, $\eta_S$ and $\xi$ are the Langevin forces for the optical and acoustic modes.   For consistency with thermodynamics, these losses and Langevin forces return the perturbed system to thermal equilibrium in the absence of electrostrictive coupling and external driving, and are consistent with the fluctuation dissipation theorem in the limit of validity for the RWA. The external decay rate $\gamma_{ext}$ accounts for the optical losses that occur when the laser resonator is coupled to a bus waveguide and quantifies the fraction of supplied pump power that can excite the resonator. The time-dependent complex amplitude $F_{ext}$ accounts for the effects of the pump laser, including noise, and is normalized so that the on-chip pump laser power can be represented by $P_p=\hbar\omega_p |F_{ext}|^2$.

The Langevin forces $\eta_p$, $\eta_S$ and $\xi$ capture the quantum and thermal fluctuations of the optical and acoustic fields, respectively, and can be modeled by zero-mean Gaussian variables with white power spectra \cite{li2012characterization,loh2015noise,behunin2018fundamental}. The two-time correlation properties for these force are given by
\begin{equation}\label{lang1}
\langle\eta_{m}^{\dagger}(t) \eta_{m^{\prime}}(t^{\prime})\rangle=\gamma_{m} N_{th,m} \delta\left(t-t^{\prime}\right) \delta_{m m'}
\end{equation}
\begin{equation}\label{lang2}
\langle\eta_{m}(t) \eta_{m^{\prime}}^{\dagger}(t^{\prime})\rangle=\gamma_{m} (N_{th,m}+1) \delta\left(t-t^{\prime}\right) \delta_{m m'}
\end{equation}
\begin{equation}\label{lang3}
\langle\xi^{\dagger}(t) \xi(t^{\prime})\rangle=\Gamma n_{th} \delta\left(t-t^{\prime}\right)
\end{equation}
\begin{equation}\label{lang4}
\langle\xi(t) \xi^{\dagger}(t^{\prime})\rangle=\Gamma (n_{th}+1) \delta\left(t-t^{\prime}\right) 
\end{equation}
where ${N_{th,m}=({\rm exp}(\hbar\omega_m/k_B T_0)-1)^{-1}}$ and ${n_{th}=({\rm exp}(\hbar\Omega/k_B T_0)-1)^{-1}}$ are the thermal occupation numbers of the optical and acoustic modes, respectively, $T_0$ is the temperature, and $\hbar$ and $k_B$ are the Planck and Boltzmann constants.

A further simplification of these dynamics can be obtained when the phonon decay rate far exceeds the optical decay rates $(\Gamma \gg \gamma_m)$. Under these conditions, the phonon field amplitude can be obtained in the quasistatic limit where the amplitudes $a_p$ and $a_S$ evolve slowly compared to $\Gamma$, giving  
\begin{equation}\label{bapprox}
    b \approx \hat{b} - i \frac{2 g^*}{\Gamma}  a_p a_S^\dagger
\end{equation}
where $\hat{b}$ quantifies the thermal and quantum fluctuations of the phonon field
\begin{equation}
    \hat{b}=\int^t_{-\infty}d\tau \ e^{-\frac{\Gamma}{2}(t-\tau)}\xi(\tau).
\end{equation}
In Appendix A, we show that the leading order correction to Eq. \eqref{bapprox} captures the frequency noise transferred from the pump laser.

Inserting Eq. \eqref{bapprox} into the Eqs. \eqref{apeom} and \eqref{aseom}, we find the effective SBS laser equations of motion 
\begin{equation}\label{apeom2}
    \dot{a}_p = -\frac{1}{2}\gamma_p a_p - \mu a_S^\dagger a_S a_p + \sqrt{\gamma_{ext}}F_{ext} + h_p 
\end{equation}
\begin{equation}\label{aseom2}
    \dot{a}_S = -\frac{1}{2}\gamma_S a_S + \mu a_p^\dagger a_p a_S + h_S 
\end{equation}
\newline
where $\mu = 2|g|^2/\Gamma$, quantifying the Brillouin coupling, is proportional to the Brillouin gain, and $h_p=\eta_p -i g a_S \hat{b}$ and $ h_S=\eta_S -i g^* a_p \hat{b}^\dagger$ are ``phonon-dressed" Langevin forces. These forces describe how quantum and thermal fluctuations of the optical and mechanical modes impart colored multiplicative noise on the optical fields through electrostriction. While these approximations reproduce much of the key physics of Brillouin lasers, one must retain the first order corrections to the quasi-static approximation taken above in Eq. \eqref{bapprox} to recover the line-narrowing properties of Brillouin lasers described by Debut {\it et al.} (see Appendix A) \cite{debut2000linewidth}.

%%%%%%%%%%%%%%%%%%%%%%%%%%%%%%%%%%%%%%%%%%%%%%%
\subsubsection{Fundamental and driven thermal noise}
%%%%%%%%%%%%%%%%%%%%%%%%%%%%%%%%%%%%%%%%%%%%%%%

To model thermo-optic noise, we add the stochastic heat equation to our model, accounting for fundamental and optically driven thermal fluctuations, and include the impact of these thermal fluctuations on the laser dynamics. To capture the latter, we add a zero-mean fluctuating frequency $\hat{\omega}_j$ to the laser equations of motion that is driven by the mode-volume-averaged temperature field. With these modifications Eqs. \eqref{apeom2} \& \eqref{aseom2} become
%F_{ext}$ captures the fluctuations of the pump laser and how that will relate to photothermal noise.  Eqs. \eqref{apeom2} and \eqref{aseom2} become:
\begin{equation}\label{apeom3}
    \dot{a}_p = (-i\hat{\omega}_p -\gamma_p/2) a_p - \mu a_S^\dagger a_S a_p+ \sqrt{\gamma_{ext}}F_{ext} + h_p 
\end{equation}
\begin{equation}\label{aseom3}
    \dot{a}_S = (-i\hat{\omega}_S-\gamma_S/2) a_S + \mu a_p^\dagger a_p a_S + h_S 
\end{equation}
\newline
where $\hat{\omega}_j$, derived using modal perturbation theory, is given by 
\begin{equation}
\label{d_omega}
    \hat{\omega}_j=-\frac{\omega_j}{n_g}\int d^3x \ \frac{dn}{dT} |E_0(\mathbf{x})|^2 T(t,\mathbf{x}).
\end{equation}
Here, $n_g$ is the group index, $dn/dT$ is the thermo-optic coefficient, $E_0$ is the volume normalized electric field strength in the cavity (i.e., $\int d^3x |E_0|^2 = 1$), and $T(t,\mathbf{x})$ represents temperature fluctuations about equilibrium \cite{gorodetsky2004fundamental}. 
The solution to the driven heat equation given by
\begin{equation}\label{heateom}
\begin{split}
    \rho C \dot{T}(t,\mathbf{x})-\nabla \cdot \kappa \nabla T(t,\mathbf{x}) = \zeta(t,\mathbf{x}) + \dot{q}_{opt}
\end{split}
\end{equation}
provides $T(t,\mathbf{x})$, 
where $\kappa$ is the thermal conductivity, $C$ is the specific heat capacity at constant volume, $\rho$ is the mass density, $\zeta(t,\mathbf{x})$ is a zero mean Langevin force driving fundamental thermal fluctuations according to the fluctuation-dissipation relation (see Ref. \cite{gorodetsky2004fundamental}), and $\dot{q}_{opt}$ describes the heat flux density produced by fluctuating optical power within the resonator \cite{boyd2020nonlinear}. Here, $\dot{q}_{opt}$ captures all changes in temperature produced by the presence of optical energy, e.g., absorption or electrostiction plus mechanical dissipation. Together, Eqs. \eqref{apeom3}, \eqref{aseom3}, and \eqref{heateom} provide a unified description of the key physics that determine SBS laser dynamics in non-crystalline media (generalization to crystalline materials requires the inclusion of thermal expansion). A rigorous treatment of thermal expansion requires a treatment of the laser resonator's thermally driven mechanical motion as well as a modification of Eq. \eqref{d_omega} to include changes in the laser resonator brought about by thermal deformations of the waveguide geometry. These effects must be includes in systems where thermal expansion cannot be neglected.

%%%%%%%%%%%%%%%%%%%%%%%%%%%%%%%%%%%%%%%%%%%%%%%
\subsubsection{Amplitude and Phase decomposition}
%%%%%%%%%%%%%%%%%%%%%%%%%%%%%%%%%%%%%%%%%%%%%%%

To explore the noise dynamics of the SBS laser described by Eqs. \eqref{apeom3}, \eqref{aseom3}, and \eqref{heateom}, we decompose $a_p$ and $a_S$ in terms of phase and amplitude expressed by
\begin{equation}\label{ap}
    a_{p}=\left(\alpha_{p}+\delta \alpha_{p}\right) e^{i \varphi_{p}}
\end{equation}
\begin{equation}\label{as}
    a_{S}=\left(\alpha_{S}+\delta \alpha_{S}\right) e^{i \varphi_{S}}.
\end{equation}
Here $\alpha_p$ and $\alpha_S$ are time-independent, steady state amplitudes of the pump and Stokes mode,  $\delta \alpha_p$ and $\delta \alpha_S$ represent fluctuations about the steady-state amplitude, and $\varphi_p$ and $\varphi_S$ are time-dependent fluctuating phases of the optical modes. The steady-state amplitudes for the pump and Stokes modes (for a single order Brillouin laser above threshold) are given by (see Ref. \cite{behunin2018fundamental})
\begin{align}
\label{ssamps}
    \alpha_p^2 & =  \frac{\gamma_S}{2\mu}\\
    \alpha_S^2 & =  \frac{1}{\mu}\bigg[ \frac{\sqrt{\gamma_{ext}} |F_{ext}|}{\alpha_p} -\frac{\gamma_p}{2}\bigg].
\end{align}

We obtain the amplitude and phase dynamics by inserting Eqs. \eqref{ap} and \eqref{as} into Eqs. \eqref{apeom3} and \eqref{aseom3}, linearize to first order in fluctuating amplitude $\delta\alpha$ (assuming $\alpha \gg \delta \alpha$), and isolate real and imaginary parts, yielding equations for the phase and amplitude of each mode      
\begin{eqnarray}\label{dphip} 
   && \dot{\varphi}_p = -\hat{\omega}_p + \frac{1}{\alpha_p}{\rm Im}[\tilde{h}_p] + \frac{\sqrt{\gamma_{ext}}}{\alpha_p}{\rm Im}[\tilde{F}_{ext}]
    \\
\label{dphis}
  &&  \dot{\varphi}_S = -\hat{\omega}_S + \frac{1}{\alpha_S}{\rm Im}[\tilde{h}_S]
    \\
\label{dalphap}
  &&  \delta\dot{\alpha}_p = -\frac{\sqrt{\gamma_{ext}}|F_{ext}|}{\alpha_p}\delta\alpha_p-2\mu\alpha_S\alpha_p\delta\alpha_S+{\rm Re}[\tilde{h}_p]\nonumber\\
    && \quad \quad \quad \quad \quad \quad +\sqrt{\gamma_{ext}}({\rm Re}[\tilde{F}_{ext}]-|F_{ext}|)
    \\
\label{dalphas}
   && \delta\dot{\alpha}_S=2\mu\alpha_p\alpha_S\delta\alpha_p+{\rm Re}[\tilde{h}_S]
\end{eqnarray}
where $\tilde{h}_p=h_p {\rm exp}(-i \varphi_p)$, $\tilde{h}_S=h_S {\rm exp}(-i \varphi_S)$, and $\tilde{F}_{ext}=F_{ext} {\rm exp}(-i\varphi_p)$.  
Because the coupling parameter $\mu$ is real when perfect phase matching is satisfied, the phase and amplitude dynamics decouple and can be analyzed independently. While Eq. \eqref{dphip} describes the free running dynamics of the phase of the optical mode driven by the pump laser $\varphi_p$, in practice the pump laser is locked to the SBS resonator using a feedback loop. Using a control theory model for this feedback loop, shown in Appendix B, the impact of this form of laser control can be determined. For the experimental case explored in this paper, where the SBS resonator linewidth is large compared to the pump laser linewidth (see Tab. I.), $\dot{\varphi}_p \approx -\hat{\omega}_p + \xi_{ext}$ where $\xi_{ext}$ is a Langevin force modeling phase diffusion and the linewidth of the pump laser.     

In the following sections, we use the dynamics described by Eqs. \eqref{dphip}$-$\eqref{dalphas} to model the frequency and intensity noise of a SBS laser.

\subsection{Relative Intensity Noise}

Relative intensity noise (RIN), produced by fluctuations of the amplitudes $\delta \alpha_p$ and $\delta \alpha_S$ lead to instability in the emitted laser power. Equations \eqref{dalphap} and \eqref{dalphas} show that the RIN of the Stokes mode has two sources: amplitude noise transferred from the pump laser and fundamental fluctuations inherent to Brillouin scattering \cite{loh2015noise}.  Even for an ideal pump laser (i.e., power stable), spontaneous Brillouin scattering is always present, leading to amplitude noise. 

%We define the two-sided power spectral density (PSD) of the RIN of the Stokes SBS by
%\begin{equation}\label{SRINP}
 %   S_s^{RIN}[\omega]=\frac{1}{P_S^2}\int_{-\infty}^{\infty}d\tau e^{i\omega\tau} \langle \delta P_S(t) \delta P_S(t-\tau)\rangle
%\end{equation}
%where $\delta P_S$ represents time-dependent deviations of the Stokes laser power from steady-state $P_S$. 

The RIN can be expressed as an amplitude power spectral density (PSD) by expressing the relative power fluctuations in terms of the amplitude fluctuations. Using Eqs. \eqref{dalphap} and \eqref{dalphas} and  $\delta P_S = 2 P_S \delta \alpha_S/\alpha_S$ \cite{behunin2018fundamental} we find   
\begin{equation}\label{SRINdalpha}
    S_S^{RIN}[\omega]=\frac{4}{\alpha_S^2}\int_{-\infty}^{\infty}d\tau \
    e^{i\omega\tau} \langle\delta\alpha_S(t+\tau)\delta\alpha_S(t)\rangle.
\end{equation}
To obtain an explicit expression for \eqref{SRINdalpha}, we solve Eqs. \eqref{dalphap} and \eqref{dalphas} in the Fourier domain, and use the substitutions 
$\Gamma_{R}=\sqrt{\gamma_{ext}}|F_{ext}|/\alpha_p$, 
$\Omega_{RIN}=2\mu \alpha_p\alpha_S$ and $\delta f=\sqrt{\gamma_{ext}}\big( {\rm Re}[\tilde{F}_{ext}]-|F_{ext}| \big)$, the latter representing amplitude fluctuations of the pump laser, to yield
\begin{equation}\label{dalphas2}
\begin{split}
    &\delta\tilde{\alpha}_S[\omega]=\\
    &\frac{\Omega_{RIN}({\rm Re}[\tilde{h}_p[\omega]]+\delta f[\omega])+(-i\omega+\Gamma_{R}){\rm Re}[\tilde{h}_S[\omega]]}{-\omega^2-i\omega\Gamma_{R}+\Omega_{RIN}^2}.
\end{split}
\end{equation}
When Eq. \eqref{dalphas2} is inserted into Eq. \eqref{SRINdalpha}, the first order Stokes SBS laser RIN is described by the summation of two PSDs (assuming no cross-correlations i.e., pump power fluctuations are independent of fundamental fluctuations $\langle\tilde{h}_S[\omega]\delta f[\omega']\rangle=0$) given by
\begin{equation}\label{SRINtot}
    S_S^{RIN}[\omega]=S_{fund}^{RIN}[\omega]+S_{trans}^{RIN}[\omega]
\end{equation}
where $S_{trans}^{RIN}\propto\langle\delta f[\omega]\delta f[\omega']\rangle$ and $S_{fund}^{RIN}$ contains correlation terms of ${\rm Re}[\tilde{h}_p[\omega]]$, ${\rm Re}[\tilde{h}_S[\omega]]$, and cross-correlation between the two. These two terms represent intensity noise produced spontaneous scattering inherent (fundamental) to Brillouin lasing and by instability in the pump laser power (transferred). 

In agreement with prior work (see Refs. \cite{behunin2018fundamental} and  \cite{loh2015noise}), we find the (single-sided) fundamental ($S_{fund}^{RIN}$) contribution to the RIN given by
\begin{widetext}
\begin{equation}\label{SRINfund}
\begin{split}
    S_{fund}^{RIN}[\omega]=\frac{8}{\alpha_S^2}\frac{1}{\Gamma_R^2 \omega^2+(\omega^2-\Omega_{RIN}^2)^2}\Big\{ & \frac{|g|^2 (1+2 n_{th})\Gamma\gamma_S(\gamma_p-\Gamma_R)^2)}{2\mu(\Gamma^2+4\omega^2)}
      +\frac{1}{4}(1+2N_{th})[\gamma_S(\Gamma_R^2+\omega^2)+\gamma_p\Omega_{RIN}^2]\Big\}
\end{split}
\end{equation}
\end{widetext} 
exhibiting a relaxation oscillation peak near $\Omega _{RIN}$. To connect with experiment we express the pump-transferred contribution to the intensity noise in terms of the (measurable) pump RIN ($S_{ext}^{RIN}$). To linear order in fluctuating terms,  $S_{trans}^{RIN}$ can be expressed as 
\begin{equation}\label{SRINtrans2}
\begin{split}
    S_{trans}^{RIN}[\omega]= \frac{\Omega_{RIN}^2}{\alpha_S^2|\Omega_{RIN}^2-i\omega\Gamma_{R}-\omega^2|^2}
     \frac{\gamma_{ext}P_{ext}}{\hbar\omega_{ext}}S_{ext}^{RIN}[\omega].
\end{split}
\end{equation}
Here, $S_{trans}^{RIN}[\omega]$ and $S_{ext}^{RIN}[\omega]$ are both single-sided PSDs. 

In addition to the PSD describing the intensity fluctuations of the SBS laser, fluctuations of the {\it total intracavity power} are critical to calculate the photothermal noise (Sec. III.B.3). The total intracavity power $P_{tot}$ and fluctuations in the total power $\delta P_{tot}$ are given by 
\begin{eqnarray}\label{totpower}
&& P_{tot} = \frac{\hbar\omega_0 v_g}{L}\Big[\alpha_p^2+\alpha_S^2\Big]
 \\
 \label{totdpower}
 &&   \delta P_{tot} \approx \frac{\hbar\omega_0 v_g}{L}\Big[ 2\alpha_p\delta\alpha_p+2\alpha_S\delta\alpha_S\Big]
\end{eqnarray}
where second-order terms in fluctuating amplitudes have been neglected. The PSD for the total RIN, relevant to photothermal noise calculated in Sec. IIIB3, is given by
\begin{equation}\label{TotRIN}
\begin{split}
 S^{RIN}_{tot}[\omega]=&\frac{1}{P_{tot}^2}\int^{\infty}_{-\infty}d\tau e^{i\omega \tau} \langle \delta P_{tot}(t + \tau) \delta P _{tot}\rangle.
  % S^{RIN}_{tot}=&\frac{4}{(\alpha_p^2+\alpha_S^2)^2}\int^{\infty}_{-\infty}d\tau e^{i\omega \tau} \langle[\alpha_p\delta\alpha_p(t+\tau)+\alpha_S\delta\alpha_S(t+\tau)] [\alpha_p\delta\alpha_p(t)+\alpha_S\delta\alpha_S(t)]\rangle
\end{split}
\end{equation}
This equation accounts for the transferred and fundamental sources of RIN for both the pump mode and the Stokes mode as well as cross-correlations between pump and Stokes amplitude fluctuations (i.e., $\langle \delta \alpha_p \delta \alpha_S \rangle \neq 0$) \cite{behunin2018fundamental}.
The expression for $S^{RIN}_{tot}$ is given in Appendix E.

%%%%%%%%%%%%%%%%%%%%%%%%%%%%%%%%%%%%%%%%%%%%%%%%%%%%%%
\subsection{Brillouin laser frequency noise}
%%%%%%%%%%%%%%%%%%%%%%%%%%%%%%%%%%%%%%%%%%%%%%%%%%%%%%
To model the frequency stability of non-crystalline chip-integrated SBS lasers, we analyze the four dominant sources of frequency noise: fundamental noise intrinsic to the physics of SBS lasing, transferred frequency noise from the pump laser, photothermal noise, and thermorefractive noise. Owing to the distinct physical origins of each of these effects, we assume that these noise sources are uncorrelated and that the power spectrum is given by the sum of PSDs for each of these processes
\begin{equation}\label{SFtot}
    S_f[\omega]=S_f^{fund}[\omega]+S_f^{TR}[\omega]+S_f^{PT}[\omega] +S_f^{trans}[\omega]
\end{equation}
where
 $S_f^{fund}[\omega]$, 
 $S_f^{TR}[\omega]$,
 $S_f^{PT}[\omega]$, 
 and 
 $S_f^{trans}[\omega]$ respectively denote the PSD for the  fundamental, thermorefractive, photothermal, and transferred noise.  In the following subsections, we characterize the PSD for each of these noise sources. The fundamental and transferred pump frequency noise can be derived directly from our coupled-mode model, whereas the thermorefractive and photothermal noise require an analysis of the heat equation that depends on the geometry of the resonator. 

\subsubsection{Fundamental SBS noise}
The fundamental noise can be derived using Eq. \eqref{dphis} when $\hat{\omega}_S$ (thermal and power driven frequency fluctuations) is neglected, yielding a Schawlow-Townes-like linedwidth whose frequency PSD is described by
\begin{equation}\label{SFfund}
  \begin{split}
    S_{f}^{fund}[\omega]=&\frac{\gamma_S}{4\pi^2 \alpha_S^2}
    \bigg[(N_{th}+1/2)+ \frac{\Gamma^2/4 }{\omega^2+\Gamma^2/4} (n_{th}+1/2) \bigg]
  \end{split}
\end{equation}
%Here, $N_{th}$ and $n_{th}$ are the thermal occupation numbers of the optical and acoustic modes, respectively, $P_s$ is the emitted power of the Stokes SBS mode, $Q_{ext}$ is the external Q factor of the resonator, and $Q_L$ is the loaded Q factor of the resonator. 
The derivation of this result is described in previous works \cite{behunin2018fundamental,gundavarapu2019sub,lee2012chemically,suh2017phonon,li2013microwave}. Being intrinsic to Brillouin lasing, the fundamental noise sets the ultimate performance limits of an SBS laser.  

%%%%%%%%%%%%%%%%%%%%%%%%%%%%%%%%%%%%%%%%%%%%%%%%%%%%%%%%
\subsubsection{Transferred frequency Noise}
%%%%%%%%%%%%%%%%%%%%%%%%%%%%%%%%%%%%%%%%%%%%%%%%%%%%%%%%
While the SBS emission linewidth can be several orders of magnitude smaller than the pump laser linewidth, the noise transferred from the pump can be significant. To capture the effect of transferred noise, adiabatic elimination of the phonon modes must be relaxed. Keeping first order corrections (as outlined in Appendix A), we find 
\begin{equation}\label{SFtrans}
  \begin{split}
    S_f^{trans} \approx \frac{\gamma_S^2}{\Gamma^2} S_f^{ext}
  \end{split}
\end{equation}
in agreement with Ref. \cite{debut2000linewidth} in the low-frequency limit.  $S_f^{ext}[\omega]$ is the frequency noise of the external pump.
%%%%%%%%%%%%%%%%%%%%%%%%%%%%%%%%%%%%%%%%%%%%%%%%%%%%%%%%
\subsubsection{Thermorefractive and photothermal noise}
%%%%%%%%%%%%%%%%%%%%%%%%%%%%%%%%%%%%%%%%%%%%%%%%%%%%%%%%
At the scale of integrated photonics, certain noise limits arise due to the size of the system that are not present in larger scale designs. At such small system volumes, thermal fluctuations become more acute, perturbing the frequencies of a resonator through the couplings between temperature and optical properties. In non-crystalline systems there are two main sources of thermally driven frequency instability to be considered: thermorefractive and photothermal noise.  Thermorefractive noise is caused by intrinsic thermodynamic fluctuations of the temperature within a resonator that perturb the refractive index, while photothermal noise is produced by refractive index changes driven by absorption of fluctuating optical field. Other noise sources can be significant in microphotonic systems as well, examples include thermoelastic, Brownian, and pondermotive noise. While these latter effects can be important in crystalline systems, they are negligible in comparison to thermorefractive and photothermal effects in the systems considered here. 

We derive the effects of thermal fluctuations on laser frequency by solving the stochastic heat equation \cite{gorodetsky2004fundamental,matsko2007whispering}. The foundation of this approach is derived from the thermodynamic relation between the zero-mean temperature fluctuations $T$ and the volume of the resonator $V$
\begin{equation}\label{Tsquared}
    \langle T^2 \rangle=\frac{k_B T_0^2}{\rho C V}.
\end{equation}

We model fluctuations in the thermal field using Eq. \eqref{heateom} \cite{gorodetsky2004fundamental,braginsky1999thermodynamical,braginsky2000thermo}, describing  thermorefractive and, for the first time to our knowledge, RIN-driven photothermal sources of frequency noise. To approximate the impact of the optical power on the thermal field, we express $\dot{q}_{opt}$ in terms of the fluctuations in the absorbed power within the resonator given by 
\begin{equation}
   \dot{q}_{opt} \approx \alpha_{\rm abs} L |{\bf E}_0({\bf x})|^2 \delta P_{tot}(t)%-n_g z/c)
\end{equation}
where $\alpha_{\rm abs}$ is the spatial decay rate of the optical modes produced by {\it absorption}, $L$ is the resonator length, and $\delta P_{tot}(t)$ is the fluctuations of optical power within the resonator \cite{boyd2020nonlinear}. In the latter, we neglect the effect of group delay, valid when the round trip time is short compared to the characteristic time changes in the power.
Additionally, this particular form for the absorbed power assumes that the spatial distribution of the optical intensity over the waveguide cross-section (but not the overall power) does not change for translations along the waveguide and neglects the impact of changes in temperature brought about through electrostrictively driven mechanical dissipation or thermal expansion. 

Following Ref. \cite{gorodetsky2004fundamental} and generalizing to nonhomogeneous systems, i.e., appropriate for waveguides comprised of multiple materials, $\zeta$ has the correlation properties given by
\begin{equation}\label{XiCorr}
%\begin{split}
    \langle\zeta(t,\mathbf{x})\zeta(t',\mathbf{x'})\rangle = 
    -2k_B T_0^2 \nabla \cdot \kappa \nabla \delta^3(\mathbf{x}-\mathbf{x}')\delta(t-t')
%\end{split}
\end{equation}
where $T_0$ is the equilibrium temperature, and $\nabla \cdot \kappa \nabla$ acts on ${\bf x}$ in the delta function. While these correlation properties are not derived from first principles, they are consistent with the fluctuation-dissipation theorem \cite{callen1951irreversibility,gorodetsky2004fundamental}, ensuring consistency with thermodynamics, and correctly reproducing Eq. \eqref{Tsquared}.  By normalizing to steady-state power, the correlation properties of the total intracavity power $\delta P_{tot}$ is described by Eqs. \eqref{SRINfund}, \eqref{SRINtrans2} and \eqref{TotRIN}, including power fluctuations of both pump and Stokes modes (see Appendix E).  

Using an eigenfunction expansion, we solve Eq. \eqref{heateom}, expressing the temperature fluctuations in terms of heat `modes' that depend on the geometry and materials of the SBS resonator. The solution for the temperature field can be broken into two components $T = T_{TR} + T_{PT}$, one quantifying fundamental thermal fluctuations ($T_{TR}$, i.e. see Eq. \eqref{Tsquared}) and a second describing temperature changes brought about by absorbed power ($T_{PT}$). Formally, $T_{TR}$ and $T_{PT}$ can be expressed as
\begin{equation}\label{T}
\begin{split}
    \left.
    \begin{array}{ll}
      T_{TR}(t,\mathbf{x}) \\
      T_{PT}(t,\mathbf{x}) \\
      \end{array}
      \right\}
    =&\int^t_{-\infty}d\tau \int d^3x' \sum_{\mu} e^{-\lambda_{\mu}(t-\tau)}\varphi^*_{\mu}(\mathbf{x'})\varphi_{\mu}(\mathbf{x})\\
    & \quad \times \left\{
    \begin{array}{ll}
     \zeta(\tau,\mathbf{x}') \\
      \alpha_{\rm abs} L|\mathbf{E}_0(\mathbf{x}')|^2\delta P_{tot}(\tau) \\
\end{array} \right.
\end{split}
\end{equation}
where the eigenfunctions $\varphi_{\mu}(\mathbf{x})$ satisfy the self-adjoint eigenvalue equation
\begin{equation}\label{eigenvalue}
    \nabla\cdot\kappa\nabla\varphi_{\mu}(\mathbf{x})=-\rho C \lambda_{\mu}\varphi_{\mu}(\mathbf{x}),
\end{equation}
determining the real eigenvalues $\lambda_\mu$ when boundary conditions are applied, and satisfying orthonormality conditions given by 
\begin{equation}
\label{orthogonality}
    \int d^3 x \ \rho C \varphi_{\mu}(\mathbf{x}) \varphi^*_{\mu'}(\mathbf{x}) = \delta_{\mu\mu'}.
\end{equation}
Here, the symbol $\mu$ is a collective index labeling the eigenfunction.  

Using Eq. \eqref{d_omega} and assuming the fundamental temperature fluctuations are uncorrelated with the intracavity power fluctuations, the PSD of the frequency noise due to thermorefractive and photothermal noise can be expressed as the sum of $S^{TR}_f[\omega]$ and $S^{PT}_f[\omega]$ defined by

\begin{equation}
\begin{split}
\label{PSDthermo}
\left.
\begin{array}{ll}
      S^{TR}_f[\omega] \\
      S^{PT}_f[\omega] \\
      \end{array}
      \right\} 
=
&\frac{f_0^2}{n^2_g} \int d^3x \int d^3x' \frac{dn(\mathbf{x})}{dT}\frac{dn(\mathbf{x'})}{dT}  \\
&\times |\mathbf{E}_0(\mathbf{x})|^2|\mathbf{E}_0(\mathbf{x'})|^2 
\left\{
\begin{array}{ll}
    S_T^{TR}[\omega;\mathbf{x},\mathbf{x'}]  \\
     S_T^{PT}[\omega;\mathbf{x},\mathbf{x'}]  \\
\end{array} 
\right.
\end{split}
\end{equation}
where the two-point temperature fluctuation power spectra are given by 
\begin{equation}\label{PSDTTR}
\left.
\begin{array}{ll}
        S_T^{TR}[\omega;\mathbf{x},\mathbf{x'}] \\
        S_T^{PT}[\omega;\mathbf{x},\mathbf{x'}] \\
      \end{array}
      \! \! \right\}\!\!
  = \!\!\int^{\infty}_{-\infty}d\tau e^{i\omega\tau}
   \left\{\!\!
\begin{array}{ll}
        \langle T_{TR}(t\!+\!\tau,\mathbf{x})T_{TR}(t,\mathbf{x'})\rangle \\
        \langle T_{PT}(t\!+\!\tau,\mathbf{x})T_{PT}(t,\mathbf{x'})\rangle. \\
      \end{array}
      \right.
\end{equation}

Using Eqs. \eqref{XiCorr}, \eqref{eigenvalue}, and \eqref{orthogonality} and the definition for the intracavity RIN, we find single-sided thermorefractive frequency noise given by
\begin{equation}\label{TRfinal}
    S^{TR}_f[\omega] = 4k_B T_0^2 f_0^2 \sum_\mu \frac{\lambda_\mu}{\omega^2 +\lambda_\mu^2} |\mathcal{E}_\mu|^2
\end{equation}
and photothermal frequency noise given by
\begin{equation}\label{PTfinal}
    S^{PT}_f[\omega] = 2(\alpha_{abs} L P_{cav} f_0)^2 \left| 
    \sum_\mu \frac{\mathcal{E}_\mu \mathcal{F}^*_\mu}{-i\omega +\lambda_\mu} \right|^2 S^{RIN}_{tot}[\omega].
\end{equation}
Here, $\mathcal{E}_\mu$ and $ \mathcal{F}_\mu$ given by
\begin{eqnarray}
    \mathcal{E}_\mu & = & \int d^3 x \ \frac{1}{n_g} \frac{dn({\bf x})}{dT} |E_0({\bf x})|^2 \varphi_\mu({\bf x})
    \\
    \mathcal{F}_\mu & = & \int d^3 x \  |E_0({\bf x})|^2 \varphi_\mu({\bf x}),
\end{eqnarray}
are overlap integrals between the heat modes and the optical mode profile $|E_0({\bf x})|^2$ and 
$P_{cav}$ is the total average intracavity power (including cavity build-up).

Under the assumptions described above, the results to this point apply to arbitrary device materials and geometries.  In the next section, we use these results to predict the RIN and frequency noise of a Si$_3$N$_4$ ring resonator SBS laser \cite{gundavarapu2019sub} and compare these predictions with measured noise spectra.

\section{Theory-experiment comparison for an integrated photonic Si$_3$N$_4$ Brillouin laser}

To validate the modeling described in Sec. III, we compare the predicted noise spectra to measurements of frequency and intensity noise of a Brillouin laser created in a high-Q Si$_3$N$_4$ waveguide resonator \cite{gundavarapu2019sub}. For these predictions we use the model parameters listed in Tab. I and measured noise spectra for the pump laser. Where possible we use parameters determined by independent measurements. However, some materials and/or measured resonator properties have been selected to improve the theory experiment agreement based on a known range of values and/or measurement uncertainty. 
\begin{table}[ht]\label{table1}%
\caption{
Table of parameters.  The coupling rate $g$, acoustic decay rate $\Gamma$, and optical decay rates $\gamma_{p,S}$ are the same for the pump and Stokes mode.
}
\begin{ruledtabular}
\begin{tabular}{ccc}
\colrule
$g$             & 1.54 kHz            & Electrostrictive coupling rate    \\
$\Gamma$        & $(2\pi)150$ MHz     & Phonon decay rate                 \\
$\gamma_{ext}$  & $(2\pi)3.9$ MHz     & External optical decay rate       \\
$\gamma_{p,S}$        & $(2\pi) 6.8$ MHz    & Loaded optical decay rate         \\
$\mu$           & 5.1 mHz             & 1/2 $\times$ Bril. ampl. rate     \\
$L$             & 0.072 m             & Resonator length                  \\
$\sigma_{r}$    & $2.8 \mu$m          & Radial mode width                 \\
$\sigma_{z}$    & $0.85 \mu$m         & Vertical mode width               \\
$Q_L$           & $28.5 \times 10^6$  & Loaded quality factor             \\
$\alpha_{\rm abs}$        & 0.035 m$^{-1}$      & Absorption factor                 \\
$P_{ext}$       & 0.042 W             & Frequency Noise On-chip power     \\
$P_{ext}$       & 0.025 W             & RIN On-chip power                 \\
$P_{S}$         & 0.006 W             & Stokes power                      \\
$dn/dT$         & $0.87 \times 10^{-5}$\cite{elshaari2016thermo} & Thermo-optic coefficient\\
$\rho$          & 2300 kg m$^{-3}$    & PECVD silicon dioxide density     \\
\multirow{2}{1em}{$C$}                & \multirow{2}{7em}{1000 J(kg K)$^{-1}$} & Thermal silicon dioxide\\
                &                     & Specific heat capacity             \\
$\kappa$        & 0.00847              & Power coupling                    \\
$\nu_g$         & $2.06\times10^8$ m/s& Optical group velocity            \\
%\{g_P, g_I, g_D, g_{II}\} & \{0, 1.7 MHz, 0, 0.039 MHz^2\} &  gain parameters \\
\end{tabular}
\end{ruledtabular}
\end{table}

\begin{figure} %  figure placement: here, top, bottom, or page
   \centering
  \includegraphics[width=8.5cm]{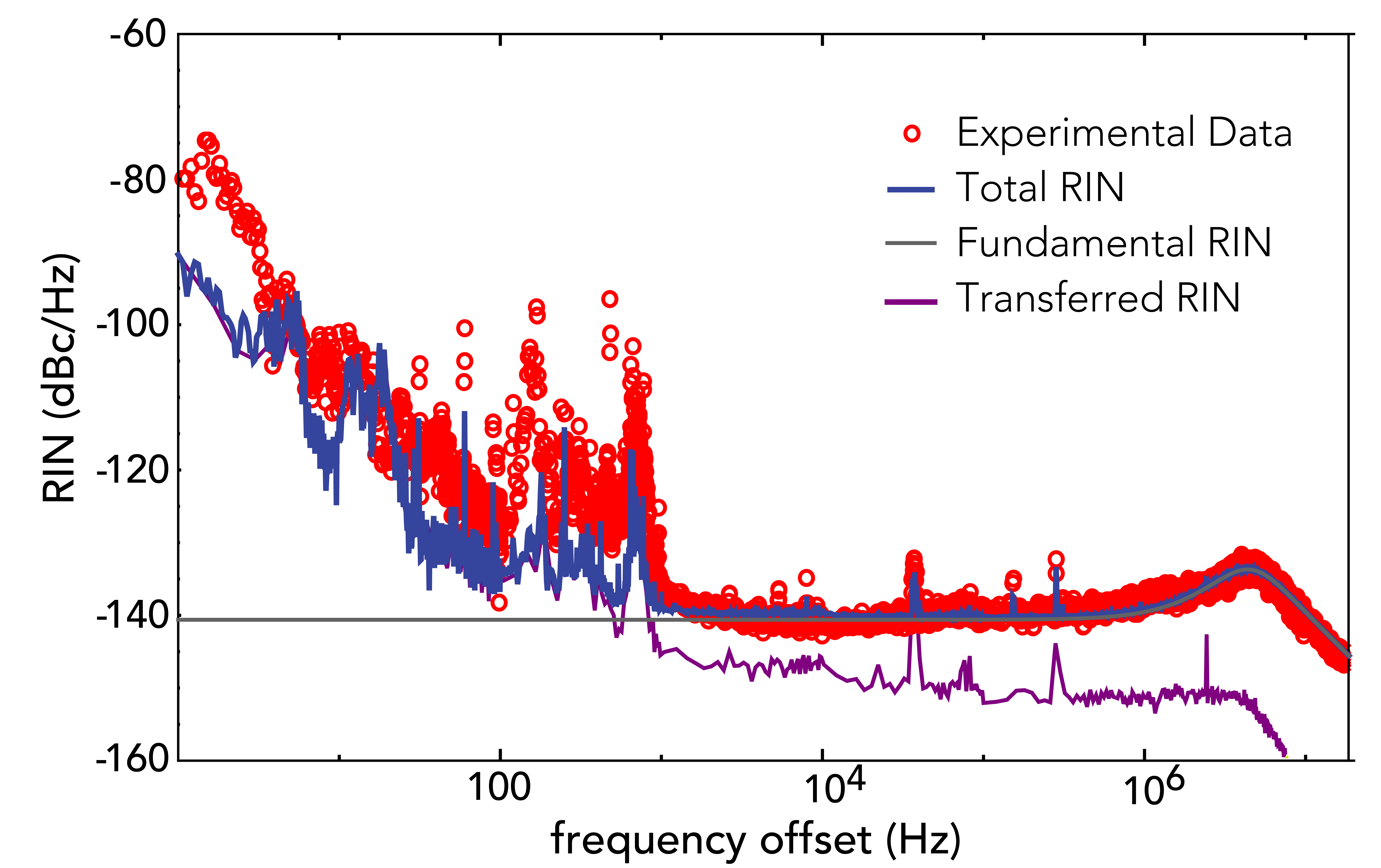}
   \caption{Relative intensity noise (RIN) of first-order, ring resonator SBS laser.  The red points are experimental data, the gray line is the fundamental RIN, the purple line is the RIN transferred from the pump, and the blue line is the composite of the models.  The analytical model includes effects due to the fundamental SBS physics and the pump laser.}
   \label{Fig: RIN total}
\end{figure}

Relative intensity noise measurements (open red circles) and predictions (blue line) are shown in Fig. \ref{Fig: RIN total}. These results show that the SBS intensity stability is well-described by transferred RIN (purple line) at low frequencies ($< 1$ kHz) and by fundamental amplitude fluctuations for Fourier frequencies above 1 kHz (gray line). A characteristic relaxation oscillation peak can be seen just above 2 MHz. 

\begin{figure} %  figure placement: here, top, bottom, or page
   \centering
   \includegraphics[width=8.5cm]{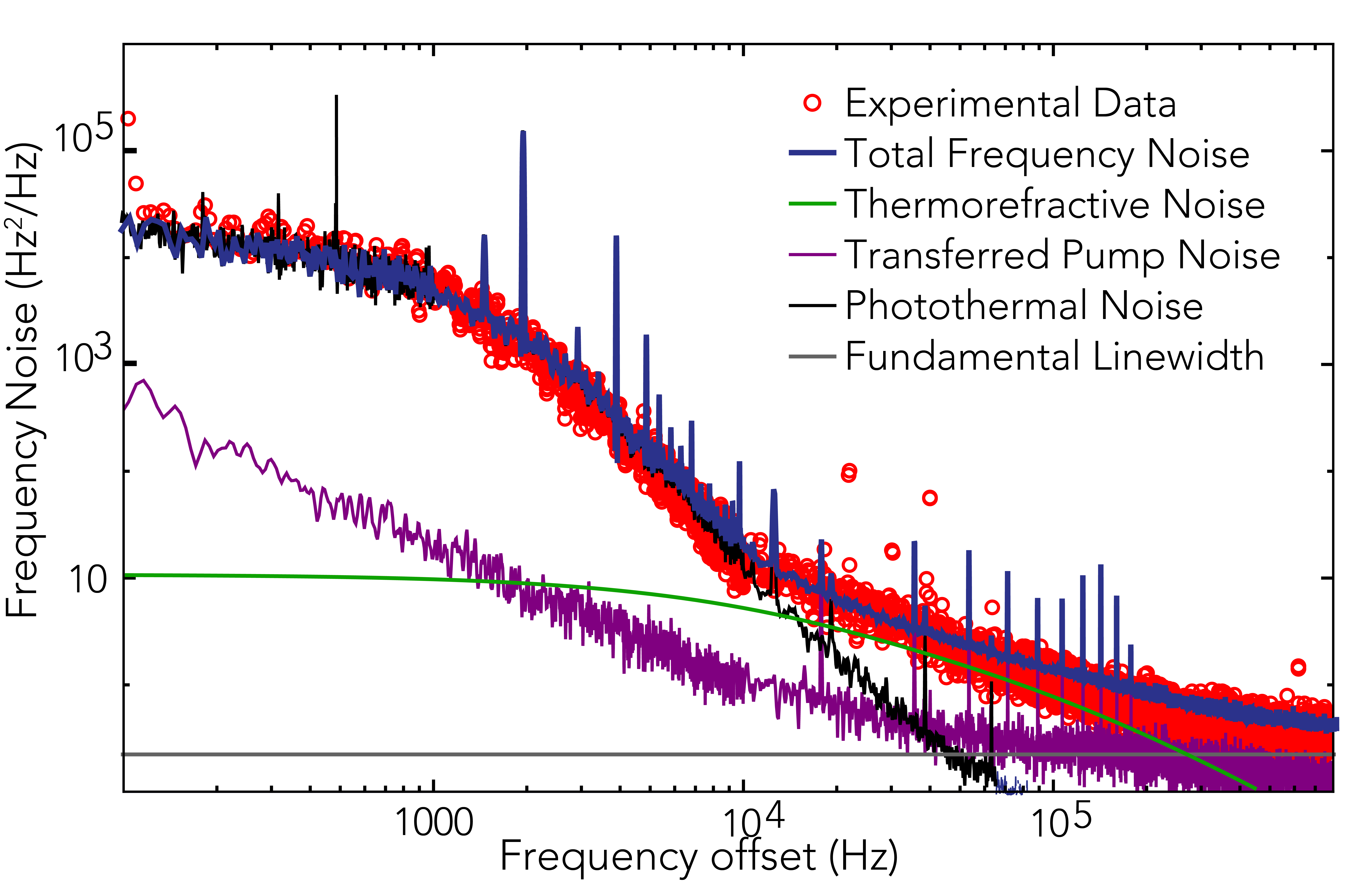}
   \caption{Frequency noise of a first-order, ring resonator SBS laser.  The experimental data is represented by the red dots. The noise sources considered in the model are: photothermal (black), transferred phase noise (purple), fundamental noise (grey), and thermorefractive (green).  The blue plot is the complete, analytical simulation of the frequency noise.  Assuming that the individual noise sources are uncorrelated, then those sources add to obtain this result.}
   \label{SBSFN}
\end{figure}

In Fig. \ref{SBSFN} predictions for fundamental and transferred frequency noise (using Eqs. \eqref{SFfund} and \eqref{SFtrans}) are directly compared with experimental data. These data show that these noise sources are small compared to thermorefractive and photothermal noise over the range of frequencies shown.

%\begin{figure} %  figure placement: here, top, bottom, or page
%   \centering
%   \includegraphics[width=8.5cm]{heat_mode_model.pdf}
%   \caption{chip-scale SBS resonator illustration}
%   \label{SBSFN}
%\end{figure}

In contrast with fundamental and transferred noise, thermal instability depends on the resonator geometry. To model these thermal effects in an Si$_3$N$_4$ ring resonator, we approximate the resonator geometry as a uniform cylinder with the same spatial volume of the actual (rectangular) chip. The waveguide forming the resonator forms a closed ring centered on the cylinder, with the radius of the SBS laser resonator, and the resonator is assumed to be comprised entirely of silica (i.e., we neglect the changes in materials property for the Si$_3$N$_4$ waveguide and the silicon handle). For a cylindrical body, the normalized eigenfunctions of the heat equation that quantify the temperature fluctuations are given by
\begin{eqnarray}\label{eigenfn}
    \varphi_{\rm m n}(r,z)=
    \frac{J_0\left(\frac{x_{0,m} r}{R}\right) \sin\left(\frac{\pi n z}{H}\right)}{\sqrt{\frac{\pi}{2}  H R^2 \rho C  J_1\left(x_{0,m}\right)^2}}
    %\sqrt{\frac{2}{\pi  H R^2 \rho C  J_1\left(x_{0,m}\right)^2}}J_0\left(\frac{x_{0,m} r}{R}\right) \sin\left(\frac{\pi n z}{H}\right)
\end{eqnarray}
where $H$ is the height of the resonator, $R$ is the radius of the cylinder, $J_{0}(x)$ is the $0$th order Bessel function of the first kind, $x_{0,m}$ is the $m$th zero of $J_0(x)$, and $n$ is an integer.  Finite element simulations show that the electromagnetic mode profile is well-approximated by the volume-normalized expression for $|E_0({r,z})|^2$ given below 
\begin{equation}\label{modelprofile}
    |E_0({r,z})|^2=\frac{1}{4 \pi^{3/2} \sigma_r \sigma_z R_c} e^{-(r-R_c)^2/\sigma_r^2-|z-H_c|/\sigma_z},
\end{equation}
where $\sigma_r$ and $\sigma_z$ are the mode widths in the radial and vertical directions, $R_c$ is the radial distance from the chip-center to the core, and $H_c$ is the vertical distance of the core from the bottom of the resonator.  We obtain the mode widths $\sigma_r$ and $\sigma_z$ by fitting \eqref{modelprofile} to simulated electromagnetic mode profile using a finite-element solver (COMSOL Multiphysics). By combining the optical mode profile Eq.\eqref{modelprofile} and the heat modes given by Eq. \eqref{eigenfn} with the equations for the thermal and driven noise Eqs. \eqref{TRfinal} and \eqref{PTfinal}, allows for the calculation of the thermorefractive and photothermal noises, respectively shown in Fig. \ref{SBSFN} as green and black lines. 

Figure \ref{SBSFN} shows that the Brillouin laser frequency noise power spectrum is well-described by a combination of photothermal, thermorefractive and fundamental noise. These results give insights about how the frequency stability of this laser can be improved. For frequencies below 10 kHz, photothermal noise (black curve) dominates and can be reduced by improving the RIN of the pump laser, lowering the intracavity power, or minimizing the optical absorption among other possibilities. Above 10 kHz, the frequency stability is determined by thermorefractive noise which can be improved by increasing the mode volume and utilizing materials with small thermo-optic coefficients.   

\section{Discussion}

In this paper we have presented a unified coupled-mode description of the dominant forms of noise for a wide array Brillouin lasers, capturing the noise transferred from the pump laser and thermally driven forms of frequency instability. In particular, we developed a model for photothermal noise in Brillouin lasers, accounting for thermal dynamics described by the heat equation. We validate our model by comparing predicted noise spectra with measurements of frequency and intensity noise for an integrated photonic Brillouin laser. Using empirically derived parameters and realistic materials properties as inputs, our model reproduces the key features of the measured noise spectra (see Figs. \ref{Fig: RIN total} \& \ref{SBSFN}) and enables the noise over a wide spectrum of frequencies to be identified and understood. 

By identifying important noise sources, our model provides insights that can pave the way to improved performance. For example, our results show that the close-to-carrier (ctc) frequency instability for an integrated photonic Si$_3$N$_4$ resonator is dominated by photothermal noise. This noise can be reduced in two ways; (1) reducing the pump laser RIN, which dominates the generation of photothermal noise at low frequencies, may drastically reduce the  ctc frequency noise, and (2), counter-intuitively, the frequency noise can be reduced by stabilizing the laser probing a second identical SBS laser resonator as reference cavity at low power. Indeed, a recent paper showed that the latter approach achieves an order of magnitude reduction in the laser linewidth and two orders of magnitude reduction in the ctc noise \cite{liu2021photonic}. As applications of ultra stable lasers transition to the chip-scale, this model provides a powerful tool to test new design concepts, diagnose noise sources, and identify paths to improved performance. 

\section{Acknowledgements}
This material is supported by Defense Advanced Research Projects Agency (FA9453-19-C-0030) and Advanced Research Projects Agency-Energy (DE-AR0001042). The views and conclusions contained in this document are those of the authors and should not be interpreted as representing official policies of DARPA, ARPA-E or the U.S. Government or any agency thereof.
\appendix

%%%%%%%%%%%%%%%%%%%%%%%%%%%%%%%%%%%%%%%%%%%%
%%%%%%%%%%%%%%%%%%%%%%%%%%%%%%%%%%%%%%%%%%%%
\section{Corrections to the quasi-static phonon approximation: pump linewidth compression}

To capture the effect of the pump noise transferred to the SBS laser, corrections to the quasi-static approximation for the phonon mode dynamics much be retained. These corrections can be derived from the exact solution to Eq. \eqref{beom} given by
\begin{equation}\label{bcomp}
    b(t)=-i g^* \int^\infty_0 dt_1 e^{-\Gamma t_1/2}a_p(t-t_1)a_S^\dagger(t-t_1)+\hat{b}.
\end{equation}
Substituting this expression for $b(t)$ into Eqs. \eqref{apeom} and \eqref{aseom}, we obtain the effective equations of motion for the Stokes mode given by
\begin{equation}\label{asalt}
  \begin{split}
    \dot{a}_S&=(-i\hat{\omega}_S-\gamma_S/2)a_S+\hat{h}_S \\
    &-|g|^2\int^\infty_0 dt_1 e^{-\Gamma t_1/2}a_S(t-t_1)a_p(t)a_p^\dagger(t-t_1)
  \end{split}
\end{equation}
Assuming that the pump and Stokes modes change very slowly in comparison to the phonon decay rate, we can approximate the the impact of the phonons on the Stokes mode by using a Markov approximation, implying
\begin{eqnarray}
  \int^\infty_0 && dt_1 e^{-\Gamma t_1/2}a_S(t-t_1)a_p(t)a_p^\dagger(t-t_1) \approx
  \\
  &&  \frac{2}{\Gamma}a_S(t)a_p(t)a_p^\dagger(t)  -\frac{4}{\Gamma^2}\bigg(\dot{a}_S(t)a_p(t)a_p^\dagger(t)\nonumber
  \\
  &&
  \quad  \quad  \quad  \quad  \quad  \quad  \quad  \quad  \quad  \quad + {a}_S(t)a_p(t)\dot{a}_p^\dagger(t)\bigg) 
  \nonumber
\end{eqnarray}
To obtain the corrections to the frequency noise as the quasi-static approximation is relaxed, we decompose Eq. \eqref{asalt} in terms of phase and amplitude using Eqs. \eqref{ap} and \eqref{as}, linearize for small perturbations of amplitude and phase about steady state, and take the imaginary part. We find
\begin{equation}\label{dphisalt}
    \begin{split}
          \dot{\varphi}_S \approx &-\hat{\omega}_S+{\rm Im}[\tilde{h}_S]-\frac{\gamma_S}{\Gamma}(-\hat{\omega}_p + \xi_{\rm ext})
    \end{split}
\end{equation}
where Eq. \eqref{ssamps}, $\mu = 2|g|^2/\Gamma$ and the results of Appendix B have been used to replace $4|g|^2 \alpha_p^2/\Gamma^2$ with $\gamma_S/\Gamma$, $\dot{\varphi}_p$ with $-\hat{\omega}_p + \xi_{\rm ext}$, and $\gamma_S/\Gamma \ll 1$ has been used. Neglecting all noise terms except for the phase diffusion of the pump laser (see Appendix B), we find the frequency noise produced by instability in the pump laser phase given by 
\begin{eqnarray}
S_f^{trans} \approx \frac{\gamma_S^2}{\Gamma^2} S_f^{ext}
\end{eqnarray}
in agreement with Debut {\it et al.} Ref. \cite{debut2000linewidth} in the limit where  $\gamma_S/\Gamma \ll 1$.

%%%%%%%%%%%%%%%%%%%%%%%%%%%%%%%%%%%%%%%%%%%%%
%%%%%%%%%%%%%%%%%%%%%%%%%%%%%%%%%%%%%%%%%%%%%

\section{Control theory and pump locking}
In practice the pump laser is locked to the resonator, modifying the laser dynamics and impacting the noise. To capture this effect, we use control theory to model the impact of this laser lock and derive equations of motion for the pump laser phase. These results show that, with realistic parameters, the carrier frequency of the pump laser tracks the resonances of the laser cavity and the pump noise is transferred to the pump light circulating in the cavity when the resonance width is much greater than pump laser linewidth.  

We model the pump laser phase $\varphi_{\rm ext}$ according to the equation of motion given by 
\begin{equation}
    \dot{\varphi}_{\rm ext} = -\omega_c + \xi_{\rm ext}
\end{equation}
where $\omega_c$ denotes the carrier frequency, that can slowly drift and be externally controlled, and $\xi_{\rm ext}$ is a $\delta$-correlated Langevin force describing the pump laser's fundamental linewidth. We assume that the pump laser carrier frequency is locked to the SBS resonator using a Pound-Drever-Hall (PDH) feedback loop  \cite{pound1946electronic,drever1983laser,black2001introduction}. Using this form of feedback, the probe laser is phase-modulated and the beatnote of the transmitted carrier and sidebands is detected. This measurement yields an error signal that quantifies the difference between the carrier frequency of the pump laser and the resonances of the optical cavity. Assuming proportional-integral-derivative (PID) feedback, a control theory model yields the equation of motion for the pump laser carrier frequency $\dot{\varphi}_{c}$ given by
\begin{eqnarray}\label{dphiext}
    \dot{\varphi}_{c}(t)=
    - g_P(\dot{\varphi}_{c}(t-\Delta t)-\dot{\hat{\varphi}}_p(t-\Delta t)) \ \
   \\
    - g_I \int_0^{t-\Delta t}d\tau \ (\dot{\varphi}_{c}(\tau)-\dot{\hat{\varphi}}_p(\tau)) \ \ \nonumber 
   \\
    - g_D(\ddot{\varphi}_{c}(t-\Delta t)-\ddot{\hat{\varphi}}_p(t-\Delta t)) \ \ \nonumber
\end{eqnarray}
where $g_P$, $g_I$ and $g_D$ are the respective proportional, integral and derivative gains of a PID controller used to stabilize the pump laser frequency \cite{zhuravlev2012development}, $\Delta t$ is the inverse loop bandwidth or loop delay, and $\hat{\varphi}_p$ (here $\dot{\hat{\varphi}}_p \equiv -\hat{\omega}_p$) represents the phase fluctuations of the resonator produced by thermorefractive and photothermal effects. 
Through the dynamics given by Eq. \eqref{dphiext}, the feedback minimizes the error signal given by $\dot{\varphi}_{c}-\dot{\hat{\varphi}}_p$, seeking to bring the pump laser in resonance with the time-dependent cavity frequency $\hat{\omega}_p$. The degree to which the pump laser is brought into resonance with the cavity is determined by loop gain and bandwidth. We assume the PDH lock operates with a modulation frequency much greater than the resonator linewidth where the gain (or frequency descriminator) is proportional to $\sqrt{P_c P_{sb}} \Delta \nu_{ext}/\Delta \nu^2$, where $P_{c}$ ($P_{sb}$) is the carrier (sideband) power sent into the cavity and $\Delta \nu$ ($\Delta \nu_{ext}$) is the loaded (external) resonator linewidth \cite{black2001introduction}. This formula generalizes the frequency discriminator for resonators with internal losses. Only for frequency fluctuations that vary slower than the loop bandwidth $1/\Delta t$ will the feedback modeled by Eq. \eqref{dphiext} effectively bring the pump laser frequency into resonance with the cavity. 

In the Fourier transform, we obtain a solution for $\varphi_{c}$ in the frequency domain given by
%\begin{equation}\label{phiext}
%    \varphi_{ext}[\omega] = \frac{\xi_{ext}[\omega] + g_I e^{i\omega\Delta %t}\hat{\varphi}_p[\omega]}{-i\omega + g_I e^{i\omega\Delta t}}.
%\end{equation}
\begin{equation}\label{phiext}
    \varphi_{c}[\omega] = \frac{D(\omega){\hat{\varphi}}_p[\omega]}{-i\omega + D(\omega)}.
\end{equation}
where $D(\omega) \equiv (-i\omega g_P + g_I-\omega^2 g_D) e^{i\omega\Delta t}$ accounts for the locking dynamics.
Next, we use this expression for $\varphi_{c}[\omega]$ to obtain phase dynamics of the mode driven by the pump laser.

Examining Eq. \eqref{dphip}, we make the substitution ${\rm Im}[\tilde{F}_{ext}]=|F_{ext}|{\rm sin}(\varphi_{ext}-\varphi_p)$ giving
%and our previous assumption of $\dot{\hat{\varphi}}_p=\hat{\omega}_p$, we get:
\begin{equation}\label{pumpeqm}
    \dot{\varphi}_p = -\hat{\omega}_p + \frac{1}{\alpha_p}{\rm Im}[\tilde{h}_p] + \frac{\sqrt{\gamma_{ext}}}{\alpha_p}|F_{ext}|\sin(\varphi_{ext}-\varphi_p)
\end{equation}
For sufficient gain in the feedback loop, i.e. {$|D(\omega)| \gg \omega$} (true within loop bandwidth for parameters in Tab. \ref{table1}), Eq. \eqref{phiext} predicts that $\varphi_{c} \approx \hat{\varphi}_p$, assuming a narrow pump laser linewidth enables a small angle approximation of $\sin(\varphi_{ext}-\varphi_p)$, so that Eq. \eqref{pumpeqm} can be solved to linear order. Under these conditions, we find 

\begin{eqnarray}
\label{pumpPhase}
    \varphi_p[\omega]
    \simeq 
    \frac{1}{i\omega}\bigg[ 1 + && \frac{i\omega \Gamma_R}{(-i\omega + \Gamma_{R})(-i\omega+D(\omega))} \bigg]\hat{\omega}_p[\omega]
        \nonumber
\\
    + && \frac{1}{-i\omega + \Gamma_{R}}\frac{1}{\alpha_p}{\rm Im}[\tilde{h}_p[\omega]] 
       \nonumber
 \\
    -
    && \frac{\Gamma_{R}}{i\omega (-i\omega +\Gamma_{R})}
    \xi_{ext}[\omega]  \quad
\end{eqnarray}
where $\Gamma_{R} \equiv \sqrt{\gamma_{ext}}|F_{ext}|/\alpha_p$. 
%\begin{eqnarray}
%\label{pumpPhase}
 %   \varphi_p[\omega] \simeq \frac{1}{-i\omega + D(\omega)+\frac{\sqrt{\gamma_{ext}}|F_{ext}|}{\alpha_p}}}
  %  \bigg[ \frac{\sqrt{\gamma_{ext}}|F_{ext}|}{\alpha_p}}\xi_{ext}[\omega]  
   % \\
    %+ \frac{1}{-i\omega}(-i\omega + D(\omega))(-\hat{\omega}_p[\omega]+ \frac{1}{\alpha_p}{\rm Im}[\tilde{h}_p[\omega]]) \nonumber
    %\bigg]  
%\end{eqnarray}
%\frac{\sqrt{\gamma_{ext}}|F_{ext}|}{\alpha_p}}
%\begin{equation}
%\begin{split}
%    \varphi_p(t)\simeq &\int^t_{-\infty}d\tau e^{\frac{-\sqrt{\gamma_{ext}}|F_{ext}|}{\alpha_p}(t-\tau)}\bigg(-\dot{\hat{\varphi}}_p(\tau) + \frac{1}{\alpha_p}{\rm Im}[\tilde{h}_p (\tau)]\\
%    &+ \frac{\sqrt{\gamma_{ext}}}{\alpha_p}|F_{ext}|\varphi_{ext}(\tau) \bigg)
%\end{split}
%\end{equation}
For the parameters listed in Tab. \ref{table1}, respective PID gains of $0, 3.4 {\rm MHz}, 0$ and loop delay $\Delta t \sim 2 \mu$s  ,  Eq. \eqref{pumpPhase} shows that the phase of the pump mode is dominated by thermorefractive and photothermal instability and the linewidth of the pump laser, permitting the approximation $\dot{\varphi}_p \approx -\hat{\omega}_p + \xi_{\rm ext}$. 

\section{Low-frequency limit of the RIN}

To examine the low frequency limit, it is convenient to manipulate the transferred RIN, Eq. \eqref{SRINtrans2}, to be:
\begin{equation}\label{SRINtransalt}
    \begin{split}
        S^{RIN}_{trans}[\omega]=&\frac{1}{|-(\omega/\Omega_{RIN})^2-i(\omega\Gamma_R/\Omega_{RIN}^2)+1|^2}\\
        &\frac{[\gamma_S/(2\mu)+\alpha_S^2]^2}{4\alpha_S^4}S^{RIN}_{ext}[\omega]
    \end{split}
\end{equation}
At the clamping point of the Stokes mode, or right as the second Stokes mode starts to lase, $\alpha_S^2=\gamma_S/(2\mu)$.  At the low frequency limit, $\omega\rightarrow0$, the transferred RIN will be identical to the pump's RIN at the clamping point.  If the second Stokes mode can be prevented from lasing, potentially by tuning the resonator to be off resonance with this mode, then an increase in the pump power will continue to decrease the Stokes RIN past the pump's RIN, further stabilizing the resonator.   

\section{Consistency of Eq. 13 and 14 with thermal equilibrium}
Here, we show that the addition of a fluctuating frequency to the resonator mode does not impact thermal equilibrium for the optical modes when (1) the Langevin force is produced by white noise, and (2) when the Langevin force and frequency noise are uncorrelated. Ordinarily, the addition of noise to a dynamical system is necessarily accompanied by additional sources of dissipation in order for finite-valued averaged quantities to exist (e.g., energy in thermal equilibrium). However, we show that when the resonator linewidth is determined by white noise, the addition of a fluctuating frequency is consistent with thermodynamics without requiring additional decay channels. In other words, zero-mean fluctuations of the resonance frequency drift do not appear to impact the average time-coincident thermodynamic properties. 

To show this, consider the dynamics of an optical mode that include the effects of fluctuating resonant frequency. In this case, the mode amplitude satisfies the Heisenberg-Langevin equation (in the rotating frame at the mean resonance frequency) given by
\begin{eqnarray}
\label{ThEqproof}
    \dot{a} = -(i\hat{\omega}(t)+\gamma/2) a + \eta.
\end{eqnarray}
Here, the dissipation rate and the Langevin force $\eta$ are selected so that thermal equilibrium is achieved at long-times, in other words that the mean mode occupation number  $\langle a^\dag(t) a(t) \rangle = N_{th}$ is given by the Bose-Einstein distribution.

Equation \eqref{ThEqproof} can be solved formally to give
\begin{eqnarray}
a(t) = \int_{-\infty}^t d\tau \ e^{-\frac{\gamma}{2}(t-\tau) -i \int_\tau^t d\tau_1 \hat{\omega}(\tau_1)} \eta(\tau).
\end{eqnarray}
Taking the expectation value of the photon number,  assuming that $\eta$ and $\hat{\omega}$ are uncorrelated, and that $\hat{\omega}$ commutes at different times, we find
\begin{eqnarray}
\langle a^\dag(t) a(t) \rangle = \int_{-\infty}^t d\tau \int_{-\infty}^t d\tau' \ e^{-\frac{\gamma}{2}(2t-\tau-\tau')}
\nonumber\\ 
\langle e^{-i\int_{\tau'}^\tau d\tau_1 \ \hat{\omega}(\tau_1)} \rangle 
\langle \eta^\dag (\tau)\eta(\tau')\rangle.
\end{eqnarray}
For white noise, i.e. $\langle \eta^\dag (\tau)\eta(\tau')\rangle = \gamma N_{th} \delta(\tau - \tau')$, we find $\langle e^{-i\int_{\tau'}^\tau d\tau_1 \ \hat{\omega}(\tau_1)} \rangle \to 1$, yielding $\langle a^\dag(t) a(t) \rangle = N_{th}$, showing that the frequency fluctuations do not impact thermal equilibrium. 

While the mean thermal occupation number is not impacted by time-dependent frequency fluctuations, the correlation properties are. For the two-time correlation function we find 
\begin{eqnarray}
\langle a^\dag(t+\tau) a(t) \rangle = N_{th} e^{-\frac{\gamma}{2}|\tau|}
\langle e^{-i\int_{t}^{t+\tau} d\tau_1 \ \hat{\omega}(\tau_1)} \rangle, 
\end{eqnarray}
where, in the special case of Gaussian frequency noise, $\langle a^\dag(t+\tau) a(t) \rangle$ becomes
\begin{eqnarray}
\label{two-timeCF}
\langle a^\dag(t+\tau) a(t) \rangle = N_{th} e^{-\frac{\gamma}{2}|\tau|}
 e^{-4\pi \int_{-\infty}^{\infty} d\omega \ \frac{\sin^2(\omega \tau)}{\omega^2} S_f[\omega] } \quad \quad
\end{eqnarray}
where $S_f[\omega]$ is the power spectrum of $\hat{\omega}/(2\pi)$. This result shows that frequency noise can alter the temporal correlations. 

\subsection{Amplitude correlations with thermorefractive noise}
Under thermal equilibrium and when the optical decay rate is much bigger than the eigenfrequency of the fundamental heat mode Eq. \eqref{two-timeCF} takes on a simple analytical form exhibiting Gaussian decay. Using Eq. \eqref{TRfinal} for $S_f$, we find
\begin{eqnarray}
 \int_{-\infty}^{\infty} d\omega \ \frac{\sin^2(\omega \tau)}{\omega^2} S_f[\omega]
=
 4 k_B T_0^2 f_0^2 \sum_\mu |\mathcal{E}_\mu|^2 \nonumber
 \\
 \times
\int_{0}^{\infty} d\omega \ \frac{\sin^2(\omega \tau)}{\omega^2}
\frac{\lambda_\mu}{\omega^2 +\lambda_\mu^2} \nonumber 
\\
\label{TR-decay}
=
 4 k_B T_0^2 f_0^2 \sum_\mu |\mathcal{E}_\mu|^2 
\frac{\pi}{4 \lambda_\mu^2}[-1 + 2 \lambda_\mu \tau  + e^{-2\lambda_\mu \tau}].
\end{eqnarray}
Here, $\gamma \gg \lambda_0$ (i.e., the fundamental heat mode frequency), enables an expansion in small $\tau$ leading to the two-time correlation function given by

\begin{eqnarray}
\label{two-timeCF2}
\langle a^\dag(t+\tau) a(t) \rangle \approx N_{th} e^{-\frac{\gamma}{2}|\tau|}
 e^{-2  \langle \delta \omega^2 \rangle \tau^2 } \quad \quad
\end{eqnarray}
where the variance in the frequency fluctuations is given by 
\begin{eqnarray}
\langle \delta \omega^2 \rangle =
 4 k_B T_0^2 (2 \pi f_0)^2 \sum_\mu |\mathcal{E}_\mu|^2
\end{eqnarray}
which can be obtained by integrating Eq. \eqref{TRfinal} over all positive $\omega$ and multiplying by $2\pi$. 

\section{Total RIN Eq. \eqref{SRINtot}}

In this appendix, we will solve Eq. (\ref{TotRIN}).  The total RIN accounts for all of the power fluctuations occurring in the resonator, including the cross-correlations between the pump and Stokes modes.  This is critical to model the photothermal noise in its entirety.   
To start, Eqs. (\ref{totpower}) and (\ref{totdpower}) are plugged into Eq. (\ref{TotRIN})
\begin{widetext}
\begin{eqnarray}\label{totRIN_2}
    S^{RIN}_{tot}=\frac{4}{(\alpha_p^2+\alpha_S^2)^2}\int^{\infty}_{-\infty}d\tau e^{i\omega \tau}\Big[&&\alpha_p^2\langle\delta\alpha_p(t+\tau)\delta\alpha_p(t)\rangle+\alpha_S^2\langle\delta\alpha_S(t+\tau)\delta\alpha_S(t)\rangle \nonumber \\
    &&+\alpha_p\alpha_S\langle\delta\alpha_p(t+\tau)\delta\alpha_S(t)\rangle+\alpha_S\alpha_p\langle\delta\alpha_S(t+\tau)\delta\alpha_p(t)\rangle\Big]
\end{eqnarray}
\end{widetext}
To solve this equation, we broke this integral up into 4 separate integrals that, when added together, gave the total RIN. Each integral can be solved using the relation $S_{\delta\alpha_i,\delta\alpha_j}[\omega]=\frac{\langle\delta\tilde{\alpha}^*_i[\omega]\delta\tilde{\alpha}_j[\omega'] \rangle}{2\pi \delta(\omega-\omega')}$, where $S_{\delta\alpha_i,\delta\alpha_j}[\omega]$ is the PSD of the amplitude fluctuations and $\delta\tilde{\alpha}_j[\omega]$ is the amplitude fluctuation represented in the Fourier domain, which is given by Eq. (\ref{dalphas2}) for the Stokes mode and can be similarly derived using Eqs. (\ref{dalphap}) and (\ref{dalphas}) for the pump mode.  Our final equation for the total RIN in resonator
\begin{widetext}
\begin{equation}
    \begin{split}
    S^{RIN}_{tot}[\omega]=\frac{4|\chi(\omega)|^2}{(\alpha_p^2+\alpha_S^2)^2}&\bigg\{\alpha^2_p\Big[\omega^2(\frac{1}{2}\gamma_S(N_{th}+\frac{1}{2})+\alpha_S^2 L_0)
    +\Omega_{RIN}^2(\frac{1}{2}\gamma_S(N_{th}+\frac{1}{2})+\alpha_p^2 L_0)
    +\omega^2\frac{\gamma_{ext}P_{ext}}{4 \hbar \omega_{ext}}S_{ext}^{RIN}[\omega]\Big]\\
    &+\alpha^2_S\Big[
    \Omega_{RIN}^2(\frac{1}{2}\gamma_S(N_{th}+\frac{1}{2})+\alpha_S^2 L_0)
    +(\omega^2+\Gamma_R^2)\left(\frac{1}{2}\gamma_S(N_{th}+\frac{1}{2})+\alpha_p^2 L_0\right)\\
    &+\Omega_{RIN}^2\frac{\gamma_{ext}P_{ext}}{4 \hbar \omega_{ext}}S_{ext}^{RIN}[\omega]-\Omega_{RIN}\Gamma_R(2\alpha_S\alpha_p L_0)\Big]-\alpha_p\alpha_S\Big[2\Omega_{RIN}\Gamma_R(\frac{1}{2}\gamma_S(N_{th}+\frac{1}{2})\\
    &+\alpha_p^2 L_0)-2\alpha_S\alpha_p L_0(\Omega_{RIN}^2-\omega^2)\Big]\bigg\}
    \end{split}
\end{equation}
where $|\chi(\omega)|^2=[(\omega^2-\Omega_{RIN}^2)^2+(\omega\Gamma_R)^2]^{-1}$ and $L_0=\frac{1}{2} g^2 (n_{th}+1/2)\frac{\Gamma}{\omega^2+\Gamma^2/4}$.
\end{widetext}
\bibliography{refs}% Produces the bibliography via BibTeX.
\end{document}